\documentclass
[superscriptaddress,secnumarabic,amssymb,amsmath,nobibnotes,aps,prd,showkeys,showpacs,nofootinbib,onecolumn,12pt]{revtex4}%
\usepackage{graphicx}
\usepackage{epsf}
\usepackage{bm}
\usepackage{amsmath}
\usepackage{amsfonts}
\usepackage{amssymb}%
\providecommand{\U}[1]{\protect\rule{.1in}{.1in}}

\newcommand{\be}{\begin{equation}}
\newcommand{\ee}{\end{equation}}

\newcommand{\mincir}{\raise
-3.truept\hbox{\rlap{\hbox{$\sim$}}\raise4.truept\hbox{$<$}\ }}
\newcommand{\magcir}{\raise
-3.truept\hbox{\rlap{\hbox{$\sim$}}\raise4.truept\hbox{$>$}\ }}

\begin{document}
\title{New conservation laws and exact cosmological solutions in Brans-Dicke cosmology with an extra scalar field}
\author{Antonios Mitsopoulos }
\email{antmits@phys.uoa.gr }
\affiliation{Faculty of Physics, Department of Astronomy-Astrophysics-Mechanics,
\ University of Athens, Panepistemiopolis, Athens 157 83, Greece}
\author{Michael Tsamparlis}
\email{mtsampa@uoa.phys.gr}
\affiliation{Faculty of Physics, Department of Astronomy-Astrophysics-Mechanics,
\ University of Athens, Panepistemiopolis, Athens 157 83, Greece}
\author{Genly Leon}
\email{genly.leon@ucn.cl}
\affiliation{Departamento de Matem\'{a}ticas, Universidad Cat\'{o}lica del Norte, Avda. Angamos 0610, Casilla 1280 Antofagasta, Chile}
\author{Andronikos Paliathanasis}
\email{anpaliat@phys.uoa.gr}
\affiliation{Institute of Systems Science, Durban University of Technology, Durban 4000,
South Africa}
\affiliation{Instituto de Ciencias F\'{\i}sicas y Matem\'{a}ticas, Universidad Austral de Chile, Valdivia 5090000, Chile}

\begin{abstract}
The derivation of conservation laws and invariant functions is an essential procedure for the investigation of nonlinear dynamical systems. In this study we consider a two-field cosmological model with scalar fields defined in the Jordan frame. In particular we consider a Brans-Dicke scalar field theory and for the second scalar field we consider a quintessence scalar field minimally coupled to gravity. For this cosmological model we apply for the first time a new technique for the derivation of conservation laws without the application of variational symmetries. The results are applied for the derivation of new exact solutions. The stability properties of the scaling solutions are investigated and criteria for the nature of the second field according to the stability of these solutions are determined. 

\end{abstract}
\date{\today}
\maketitle

\section{Introduction}

The detailed analysis of recent cosmological observations indicates that
the universe has been through two accelerating phases \cite{n1,n2,n3,n4}. The
current acceleration era is assumed to be driven by an unknown source known as dark
energy, whose main characteristic is the negative pressure which
provides an anti-gravity effect \cite{n5}. Furthermore, the early-universe
acceleration era, known as inflation, is described by a scalar field, the
inflaton, which is used to explain the homogeneity and isotropy of the present
universe. In particular, the scalar field dominates the dynamics
and explains the expansion era \cite{Aref1,guth}. Nevertheless, the scalar
field inflationary models are mainly defined on homogeneous spacetimes, or on
background spaces with small inhomogeneities \cite{st1,st2}. In \cite{w1} it
was found that the presence of a positive cosmological constant in Bianchi
cosmologies leads to expanding Bianchi spacetimes, evolving towards the de
Sitter universe. That was the first result to support the cosmic
\textquotedblleft no-hair\textquotedblright\ conjecture \cite{nh1,nh2}. This
latter conjecture states that all expanding universes with a positive
cosmological constant admit as asymptotic solution the de Sitter universe. The
necessity of the de Sitter expansion is that it provides a rapid expansion for
the size of the universe such that the latter effectively loses its memory on
the initial conditions, which implies that the de Sitter expansion solves the
\textquotedblleft flatness\textquotedblright, \textquotedblleft
horizon\textquotedblright\ and the monopole problem \cite{f1,f2}.

In the literature scalar fields have been introduced in the gravitational
theory in various ways. The simplest scalar field model is the quintessence
model, which consists of a scalar field minimally coupled to gravity
\cite{Ratra,Barrow}. Another family of scalar fields are those which belong to
the scalar-tensor theory. In this theory the scalar field is non-minimally coupled
to gravity which makes it essential for the physical state of the theory.
Another important characteristic of the scalar-tensor theories is that they
are consistent with Mach's principle. The most common scalar-tensor theory
is the Brans-Dicke theory \cite{Brans} which is considered in this study.
For other scalar-tensor theories and generalizations we refer the
reader to \cite{faraonibook,sf1,sf2,sf3,sf4,sf5,sf5a,sf5b} and references therein.

According to the cosmological principle in large scale the universe is
assumed to be homogeneous, isotropic and spatially flat. This implies that the
background space is described by the Friedmann - Lema\^{\i}tre - Robertson - Walker (FLRW) spacetime. This spacetime is characterized by the scale factor
which defines the radius of the three-dimensional (3d) Euclidean space. Since
General Relativity is a second order theory the
field equations involve second order derivatives of the scale factor. For simple
cosmological fluids like the ideal gas or the cosmological constant, the field
equations can be solved explicitly \cite{amen1}. However, when additional
degrees of freedom are introduced, like a scalar field, the field equations
cannot be solved with the use of closed-form functions and techniques of
analytic mechanics and one looks for First Integrals (FIs) which establish their (Liouville)
integrability  \cite{sym1,sym2,sym3,sym4}. The standard method for
the determination of FIs is Noether's theory \cite{sym5}.
However, there have appeared alternative geometric  methods \cite{Katzin 1973,Katzin 1981,Katzin 1982,Horwood 2007, Tsamparlis 2020, Tsamparlis 2020B} which
use the symmetries of the metric defined by the kinetic energy in order to
determine the FIs of the dynamic equations.
In the following we shall make use of one such approach in order to
determine the FIs (conservation laws) of the field equations.

In the present study we consider a cosmological model in which the gravitational Action
Integral is that of Brans-Dicke theory with an additional scalar field minimally coupled to
gravity \cite{Mukherjee2019,anbd1}. This two-scalar field model belongs to the
family of multi-scalar field models which have been used  as unified dark
energy models \cite{sf6,sf7,sf8} or as alternative models for the description
of the acceleration phases of the universe \cite{sf9,sf10,sf11,sf12}. Furthermore,
multi-scalar field models can attribute the additional degrees of
freedom provided by the alternative theories of gravity \cite{lan1,lan2,lan3}.
The structure of the paper is as follows.

In Section \ref{sec2}, we define the cosmological model
and we present the gravitational field equations. In Section \ref{sec.nonlin},
we present some important results on the derivation of quadratic first integrals (QFIs)
for a family of second order ordinary differential
equations (ODEs) with linear damping and perform a classification according to the
admitted conservation laws. The results are applied to the cosmological model
we consider in Section \ref{sec.exact.solution} where we construct
the conservation laws for the gravitational field equations. Due to the
non-linearity of the field equations it is not possible to write the general solution of
the field equation in closed-form. However, we find some exact closed-form
solutions with potential interest for the description of the cosmological
history. The stability of these exact solutions is investigated in section
\ref{sec.stab}. Finally, in Section \ref{sec.con} we summarize our results and
we draw our conclusions.

\section{Cosmological model}

\label{sec2}

For the gravitational Action Integral we consider that of Brans-Dicke scalar
field theory with an additional matter source leading to the expression
\cite{Brans,faraonibook}%
\begin{equation}
S=\int d^{4}x\sqrt{-g}\left[  \frac{1}{2}\phi R-\frac{1}{2}\frac{\omega_{BD}%
}{\phi}g^{\mu\nu}\phi_{;\mu}\phi_{;\nu}+L_{\psi}\left(  \psi,\psi_{;\mu
}\right)  \right]  +S_{m} \label{bd.01}%
\end{equation}
where $\phi\left(  x^{\kappa}\right)  $ denotes the Brans-Dicke scalar field
and $\omega_{BD}$ is the Brans-Dicke parameter. The action $S_{m}$ is assumed
to describe an ideal gas with constant equation of state parameter and the
Lagrangian function $L_{\psi}\left(  \psi,\psi_{;\mu}\right)  $ corresponds to
the second scalar field $\psi\left(  x^{\kappa}\right)  $ which is assumed to
be that of quintessence and minimally coupled to the Brans-Dicke scalar field. With these assumptions
the Action Integral (\ref{bd.01}) takes the following form%
\begin{equation}
S=\int d^{4}x\sqrt{-g}\left[  \frac{1}{2}\phi R-\frac{1}{2}\frac{\omega_{BD}%
}{\phi}g^{\mu\nu}\phi_{;\mu}\phi_{;\nu}-\frac{1}{2}g^{\mu\nu}\psi_{;\mu}%
\psi_{;\nu}-V\left(  \psi\right)  \right]  +S_{m}. \label{bd.01a}%
\end{equation}

The gravitational field equations follow from the variation of the Action
Integral (\ref{bd.01a}) with respect to the metric tensor. They are
\begin{equation}
G_{\mu\nu}=\frac{\omega_{BD}}{\phi^{2}}\left(  \phi_{;\mu}\phi_{;\nu}-\frac
{1}{2}g_{\mu\nu}g^{\kappa\lambda}\phi_{;\kappa}\phi_{;\lambda}\right)
+\frac{1}{\phi}\left(  \phi_{;\mu\nu}-g_{\mu\nu}g^{\kappa\lambda}\phi
_{;\kappa\lambda}\right)  +\frac{1}{\phi}T_{\mu\nu}\label{bd.02}%
\end{equation}
where $G_{\mu\nu}=R_{\mu\nu}-\frac{1}{2}Rg_{\mu\nu}$ is the Einstein tensor.
The energy-momentum tensor $T_{\mu\nu}={}^{\psi}T_{\mu\nu}+{}^{m}T_{\mu\nu}$
where ${}^{m}T_{\mu\nu}$ corresponds to the ideal gas and ${}^{\psi}T_{\mu\nu
}$ provides the contribution of the field $\psi\left(  x^{k}\right)  $ in the
field equations.

Concerning the equations of motion for the matter source and the two scalar
fields, we find ${}^{m}T_{\mu\nu;\sigma}g^{\mu\sigma}=0,$ while variation with
respect to the fields $\phi\left(  x^{\kappa}\right)  $ and $\psi\left(
x^{\kappa}\right)  $ provides the second order differential equations
\begin{equation}
g^{\mu\nu}\phi_{;\mu\nu} -\frac{1}{2\phi}g^{\mu\nu}\phi_{;\mu} \phi_{;\nu}
+\frac{\phi}{2\omega_{BD}}R=0 \label{bd.02a}%
\end{equation}%
\begin{equation}
g^{\mu\nu}\psi_{;\mu\nu} -\frac{dV}{d\psi} =0. \label{bd.02b}%
\end{equation}

We assume the background space to be the Friedmann - Lema\^{\i}tre -
Robertson - Walker (FLRW) spacetime with line element
\begin{equation}
ds^{2}=-dt^{2}+a^{2}(t)\left(  dx^{2}+dy^{2}+dz^{2}\right)  \label{bd.05}%
\end{equation}
where $a(t)$ is the scale factor of the universe and $H\left(  t\right)
=\frac{\dot{a}}{a}$ is the Hubble function. We note that a dot indicates
derivative with respect to the cosmic time $t$.

From the line element (\ref{bd.05}) follows that the Ricci scalar is
$R=6\left[  \frac{\ddot{a}}{a}+\left(  \frac{\dot{a}}{a}\right)  ^{2}\right]
$. Replacing in the gravitational field equations (\ref{bd.02}) we obtain
\begin{equation}
3\left(  \frac{\dot{a}}{a}\right)  ^{2}=\frac{\omega_{BD}}{2} \left(
\frac{\dot{\phi}}{\phi} \right)  ^{2} -3 \frac{\dot{a}}{a} \frac{\dot{\phi}%
}{\phi} +\frac{\rho_{m}+\rho_{\psi}}{\phi} \label{bd.10}%
\end{equation}%
\begin{equation}
2\frac{\ddot{a}}{a} +\left(  \frac{\dot{a}}{a}\right)  ^{2} = -\frac
{\omega_{BD}}{2}\left(  \frac{\dot{\phi}}{\phi}\right)  ^{2}-2 \frac{\dot{a}%
}{a} \frac{\dot{\phi}}{\phi} -\frac{\ddot{\phi}}{\phi} -\frac{p_{m}+p_{\psi}%
}{\phi} \label{bd.11}%
\end{equation}
where $\rho_{m}, p_{m}$ are the mass density and the isotropic pressure of the
ideal gas; and for the quintessence field
\begin{equation}
\rho_{\psi}=\frac{1}{2}\dot{\psi}^{2}+V\left(  \psi\right)  , \enskip p_{\psi
}=\frac{1}{2}\dot{\psi}^{2}-V\left(  \psi\right)  \text{.} \label{bd.12}%
\end{equation}

For the equations of motion for the scalar fields we find%
\begin{equation}
\ddot{\phi}+3 \frac{\dot{a}}{a} \dot{\phi} = \frac{\left(  \rho_{m}%
-3p_{m}\right)  +\left(  \rho_{\psi}-3p_{\psi}\right)  }{2\omega_{BD}+3}
\label{bd.13}%
\end{equation}
and%
\begin{equation}
\ddot{\psi}+3H\dot{\psi}+ \frac{dV}{d\psi}=0. \label{bd.14a}%
\end{equation}

Finally, for the matter source the continuity equation ${}^{m}T_{\mu\nu
;\sigma}g^{\mu\sigma}=0$ reads
\begin{equation}
\dot{\rho}_{m}+3\frac{\dot{a}}{a}\left(  \rho_{m}+p_{m}\right)  =0.
\label{bd.15}%
\end{equation}
For an ideal gas the equation of state is $p_{m}=w_{m}\rho_{m}$, where $w_{m}$ is an arbitrary
constant. Substituting in equation (\ref{bd.15}) we
find the solution
\begin{equation}
\rho_{m} =\rho_{m0} a^{-3\left(  1+w_{m}\right)  } \label{bd.16}%
\end{equation}
where $\rho_{m0}$ is an arbitrary constant.

The system of the ODEs that should be solved
consists of the differential equations (\ref{bd.10}), (\ref{bd.11}), (\ref{bd.13}) and
(\ref{bd.14a}).

\section{Quadratic first integrals for a class of second order ODEs with linear damping}

\label{sec.nonlin}

Consider the second order ODE
\begin{equation}
\ddot{x}=-\omega(t)x^{n} +\Phi(t)\dot{x} \label{eq.nonl1}%
\end{equation}
where the constant $n\neq-1$. In the following we shall determine the relation
between the functions $\omega(t), \Phi(t)$ for which the ODE (\ref{eq.nonl1})
admits a quadratic first integral (QFI). The case of linear first integrals
(LFIs) is also included in our study.

This problem has been considered previously in \cite{Da Silva1974},
\cite{Sarlet1980} (see eq. (28a) in \cite{Da Silva1974} and eq. (17) in
\cite{Sarlet1980}) and has been answered partially using different methods. In
\cite{Da Silva1974} the author used the Hamiltonian formalism where one looks
for a canonical transformation to bring the Hamiltonian in a time-separable
form. In \cite{Sarlet1980} the author used a direct method for constructing
FIs by multiplying the equation with an integrating factor. In
\cite{Sarlet1980} it is shown that both methods are equivalent and that the
results of \cite{Sarlet1980} generalize those of \cite{Da Silva1974}. In the
following we shall generalize the results of \ \cite{Sarlet1980}.

Equation (\ref{eq.nonl1}) is equivalent (see e.g. \cite{LeoTsampAndro2017}) to
the equation
\begin{equation}
\frac{d^{2}x}{d\tau^{2}}=-\bar{\omega}(\tau)x^{n},\enskip n\neq-1
\label{eq.nonl2}%
\end{equation}
where the function $\bar{\omega}(\tau)$ and the new independent variable
$\tau$ are defined as
\begin{equation}
\tau(t)=\int e^{\int\Phi(t)dt}dt~,~\bar{\omega}(\tau)=\omega(t(\tau))\left(
\frac{dt}{d\tau}\right)  ^{2}\iff\omega(t)=\bar{\omega}(\tau(t))e^{2\int
\Phi(t)dt}. \label{eq.damp0c}%
\end{equation}

We assume that equation (\ref{eq.nonl2}) admits the general quadratic
first integral
\begin{equation}
I=K_{11}(\tau,x)\left(  \frac{dx}{d\tau}\right)  ^{2}+K_{1}(\tau,x)\frac
{dx}{d\tau}+K(\tau,x) \label{eq.nonl4}%
\end{equation}
where the unknown coefficients $K,K_{1},K_{11}$ are arbitrary functions of
$\tau,x$. We impose  the condition
\begin{equation}
\frac{dI}{d\tau}=0. \label{eq.nonl2.1}%
\end{equation}
Replacing the second derivatives $\frac{d^{2}x}{d\tau^{2}}$, whenever they
appear using  equation (\ref{eq.nonl2}) we find that
the function $K_{11}=K_{11}(\tau)$ and the following system of
equations must be satisfied
\begin{align}
K_{1}(\tau,x)  &  =-\frac{dK_{11}}{d\tau}x+b_{1}(\tau)\label{eq.nonl3a}\\
K(\tau,x)  &  =2\bar{\omega}K_{11}\frac{x^{n+1}}{n+1}+\frac{d^{2}K_{11}}%
{d\tau^{2}}\frac{x^{2}}{2}-\frac{db_{1}}{d\tau}x+b_{2}(\tau) \label{eq.nonl3b}%
\\
0  &  =\left(  \frac{2\frac{d\bar{\omega}}{d\tau}K_{11}}{n+1}+\frac
{2\bar{\omega}\frac{dK_{11}}{d\tau}}{n+1}+\bar{\omega}\frac{dK_{11}}{d\tau
}\right)  x^{n+1}-\bar{\omega}b_{1}x^{n}+\frac{d^{3}K_{11}}{d\tau^{3}}%
\frac{x^{2}}{2}-\frac{d^{2}b_{1}}{d\tau^{2}}x+\frac{db_{2}}{d\tau}
\label{eq.nonl3c}%
\end{align}
where $b_{1}(\tau),b_{2}(\tau)$ are arbitrary functions.

We consider the solution of the latter system (\ref{eq.nonl3a}) -
(\ref{eq.nonl3c}) for various values of the power $n$.

As will be shown for the values $n=0,1,2$ there results a family of
`frequencies' $\bar{\omega}(\tau)$ parameterized with functions, whereas for
the values $n\neq-1$ results a family of `frequencies' $\bar{\omega}(\tau)$
parameterized with constants.

\subsection{Case $n=0$}

For $n=0$ the QFI (\ref{eq.nonl4}) becomes
\begin{equation}
I=K_{11}\left(  \frac{dx}{d\tau}\right)  ^{2} -\frac{dK_{11}}{d\tau}
x\frac{dx}{d\tau} +b_{1}(\tau)\frac{dx}{d\tau} +c_{3}x^{2} +2\bar{\omega}%
(\tau) K_{11}x-\frac{db_{1}}{d\tau}x+\int b_{1}(\tau) \bar{\omega}(\tau)
d\tau\label{eq.nonl4.1}%
\end{equation}
where $K_{11}=c_{1}+c_{2}\tau+c_{3}\tau^{2}$, the parameters $c_{1}%
,c_{2},c_{3}$ are arbitrary constants and the functions $b_{1}(\tau),
\bar{\omega}(\tau)$ satisfy the condition
\begin{equation}
\frac{d^{2}b_{1}}{d\tau^{2}} =2\frac{d\bar{\omega}}{d\tau}K_{11} +3\bar
{\omega}\frac{dK_{11}}{d\tau}. \label{eq.nonl4.2}%
\end{equation}

Using the transformation (\ref{eq.damp0c}) equations (\ref{eq.nonl4.1}),
(\ref{eq.nonl4.2}) become
\begin{align}
I  &  =\left[  c_{1}+c_{2}\int e^{\int\Phi\left(  t\right)  dt}dt+c_{3}\left(
\int e^{\int\Phi\left(  t\right)  dt}dt\right)  ^{2}\right]  e^{-2\int
\Phi\left(  t\right)  dt}\dot{x}^{2} \nonumber \\
& -\left[  c_{2}+2c_{3}\int e^{\int
\Phi\left(  t\right)  dt}dt\right]  e^{-\int\Phi\left(  t\right)  dt}x\dot
{x}   +b_{1}(\tau(t))e^{-\int\Phi\left(  t\right)  dt}\dot{x}+c_{3}x^{2}%
\nonumber\\
&+2\omega(t)\left[  c_{1}+c_{2}\int e^{\int\Phi\left(  t\right)  dt}%
dt+c_{3}\left(  \int e^{\int\Phi\left(  t\right)  dt}dt\right)  ^{2}\right]
e^{-2\int\Phi\left(  t\right)  dt}x \nonumber\\
&  -\dot{b}_{1}e^{-\int\Phi\left(  t\right)  dt}x+\int b_{1}(\tau
(t))\omega(t)e^{-\int\Phi\left(  t\right)  dt}dt \label{eq.nonl4.2.1}%
\end{align}
and
\begin{align}
\ddot{b}_{1}-\Phi\dot{b}_{1}  &  =2e^{-\int\Phi\left(  t\right)  dt}\left(
\dot{\omega}-2\Phi\omega\right)  \left[  c_{1}+c_{2}\int e^{\int\Phi\left(
t\right)  dt}dt+c_{3}\left(  \int e^{\int\Phi\left(  t\right)  dt}dt\right)
^{2}\right]   \nonumber\\
&  +3\omega\left[  c_{2}+2c_{3}\int e^{\int\Phi\left(  t\right)  dt}dt\right]
. \label{eq.nonl4.2.2}%
\end{align}

\subsection{Case $n=1$}

For $n=1$, we derive the well-known results of the one-dimensional (1d)
time-dependent oscillator (see e.g. \cite{Katzin1974,Prince1980}).
Specifically, we find for the frequency $\bar{\omega}(\tau)=-\frac{1}{b_{1}%
}\frac{d^{2}b{1}}{d\tau^{2}}$ the LFI
\begin{equation}
I_{1}=b_{1}\frac{dx}{d\tau}-\frac{db_{1}}{d\tau}x\label{osc1}%
\end{equation}
and for the frequency $\bar{\omega}(\tau)=-\frac{1}{2K_{11}}\frac{d^{2}K_{11}%
}{d\tau^{2}}+\frac{1}{4K_{11}^{2}}\left(  \frac{dK_{11}}{d\tau}\right)
^{2}+\frac{c_{0}}{2K_{11}^{2}}$, where $c_{0}$ is an arbitrary constant, the
QFI\footnote{For $K_{11}=\rho(\tau)^{2}$, where $\rho(\tau)$ is an arbitrary
function, the QFI takes the usual form of the Lewis invariant.}
\begin{equation}
I_{2}=K_{11}\left[  \left(  \frac{dx}{d\tau}\right)  ^{2}+\bar{\omega}%
x^{2}\right]  -\frac{dK_{11}}{d\tau}x\frac{dx}{d\tau}+\frac{d^{2}K_{11}}%
{d\tau^{2}}\frac{x^{2}}{2}.\label{osc2}%
\end{equation}
Using the transformation (\ref{eq.damp0c}) we deduce that the original
equation
\begin{equation}
\ddot{x}=-\omega(t)x+\Phi\left(  t\right)  \dot{x}\label{eq.nonl4.2.5}%
\end{equation}
for the frequency
\begin{equation}
\omega(t)=-\rho^{-1}\ddot{\rho}+\Phi\rho^{-1}\dot{\rho}+\rho^{-4}e^{2\int
\Phi\left(  t\right)  dt}\label{eq.nonl4.2.6}%
\end{equation}
admits the general solution
\begin{equation}
x(t)=\rho(t)\left(  A\sin\theta+B\cos\theta\right)  \label{eq.nonl4.2.7}%
\end{equation}
where $A,B$ are arbitrary constants, $\rho(t)\equiv\rho(\tau(t))$ and
$\theta(\tau(t))=\int\rho^{-2}(t)e^{\int\Phi\left(  t\right)  dt}dt$.

\subsection{Case $n=2$}

For $n=2$, we derive the function $\bar{\omega}=K_{11}^{-5/2}$ and the QFI
\begin{equation}
I=K_{11}(\tau)\left(  \frac{dx}{d\tau}\right)  ^{2}-\frac{dK_{11}}{d\tau
}x\frac{dx}{d\tau}+(c_{4}+c_{5}\tau)\frac{dx}{d\tau}+\frac{2}{3}K_{11}%
^{-3/2}x^{3}+\frac{d^{2}K_{11}}{d\tau^{2}}\frac{x^{2}}{2}-c_{5}x
\label{eq.nonl4.3}%
\end{equation}
where $c_{4},c_{5}$ are arbitrary constants and the function $K_{11}(\tau)$ is
given by
\begin{equation}
\frac{d^{3}K_{11}}{d\tau^{3}}=2(c_{4}+c_{5}\tau)K_{11}^{-5/2}.
\label{eq.nonl4.411}%
\end{equation}

Using the transformation (\ref{eq.damp0c}) the above results become
\begin{equation}
\omega(t)=K_{11}^{-5/2}e^{2\int\Phi\left(  t\right)  dt} \label{eq.nonl4.4.1}%
\end{equation}%
\begin{align}
I  &  =K_{11}e^{-2\int\Phi\left(  t\right)  dt}\dot{x}^{2}-\dot{K}%
_{11}e^{-2\int\Phi\left(  t\right)  dt}x\dot{x}+\left[  c_{4}+c_{5}\int
e^{\int\Phi\left(  t\right)  dt}dt\right]  e^{-\int\Phi\left(  t\right)
dt}\dot{x}\nonumber\\
& +\frac{2}{3}K_{11}^{-3/2}x^{3}  +\left(  \ddot{K}_{11}-\Phi\dot{K}_{11}\right)  e^{-2\int\Phi\left(
t\right)  dt}\frac{x^{2}}{2}-c_{5}x \label{eq.nonl4.4.2}%
\end{align}
and
\begin{equation}
\dddot{K}_{11}-3\Phi\ddot{K}_{11}-\dot{\Phi}\dot{K}_{11} +2\Phi^{2}\dot
{K}_{11}=2\left[  c_{4}+c_{5}\int e^{\int\Phi\left(  t\right)  dt}dt\right]
e^{3\int\Phi\left(  t\right)  dt}K_{11}^{-5/2} \label{eq.nonl4.4.3}%
\end{equation}
where now the function $K_{11}=K_{11}(\tau(t))$.

We note that for $n=2$ equation (\ref{eq.nonl1}), or to be more specific its
equivalent (\ref{eq.nonl2}), arises in the solution of Einstein field
equations when the gravitational field is spherically symmetric and the matter
source is a shear-free perfect fluid (see e.g.
\cite{StephaniB,Stephani1983,Srivastana1987,Leach1992,LeachMaartens1992,Maharaj1996}%
).

\subsection{Case $n\neq-1$}

For $n\neq-1$ we find $b_{1}=b_{2}=0$, $K_{11}=c_{1}+c_{2}%
\tau+c_{3}\tau^{2}$ and $\bar{\omega}(\tau)=(c_{1}+c_{2}\tau+c_{3}\tau
^{2})^{-\frac{n+3}{2}}$ where $c_{1},c_{2},c_{3}$ are arbitrary constants.

The QFI (\ref{eq.nonl4}) is
\begin{equation}
I=(c_{1}+c_{2}\tau+c_{3}\tau^{2})\left(  \frac{dx}{d\tau}\right)  ^{2}%
-(c_{2}+2c_{3}\tau)x\frac{dx}{d\tau}+\frac{2}{n+1}(c_{1} +c_{2}\tau+c_{3}%
\tau^{2})^{-\frac{n+1}{2}}x^{n+1}+c_{3}x^{2} \label{eq.nonl5}%
\end{equation}
and the function
\begin{equation}
\bar{\omega}(\tau)=(c_{1}+c_{2}\tau+c_{3}\tau^{2})^{-\frac{n+3}{2}}.
\label{eq.nonl6}%
\end{equation}

It has been checked that (\ref{eq.nonl5}), (\ref{eq.nonl6}) for $n=0,1,2$ give
results compatible with the ones we found for these values of $n$. Using the
transformation (\ref{eq.damp0c}) we deduce that the original system
(\ref{eq.nonl1}) is integrable iff the functions $\omega(t),~\Phi\left(
t\right)  $ are related as follows
\begin{equation}
\omega(t)=\left[  c_{1}+c_{2}\int e^{\int\Phi\left(  t\right)  dt}%
dt+c_{3}\left(  \int e^{\int\Phi\left(  t\right)  dt}dt\right)  ^{2}\right]
^{-\frac{n+3}{2}}e^{2\int\Phi\left(  t\right)  dt}. \label{eq.nonl6.1}%
\end{equation}
In this case the associated QFI (\ref{eq.nonl5}) is
\begin{align}
I  &  =\left[  c_{1}+c_{2}\int e^{\int\Phi\left(  t\right)  dt}dt+c_{3}\left(
\int e^{\int\Phi\left(  t\right)  dt}dt\right)  ^{2}\right]  e^{-2\int
\Phi\left(  t\right)  dt}\dot{x}^{2} \nonumber\\
& -\left[  c_{2}+2c_{3}\int e^{\int
\Phi\left(  t\right)  dt}dt\right]  e^{-\int\Phi\left(  t\right)  dt}x\dot
{x} \nonumber\\
&  +\frac{2}{n+1}\left[  c_{1}+c_{2}\int e^{\int\Phi\left(  t\right)
dt}dt+c_{3}\left(  \int e^{\int\Phi\left(  t\right)  dt}dt\right)
^{2}\right]  ^{-\frac{n+1}{2}}x^{n+1}+c_{3}x^{2}. \label{eq.nonl6.2}%
\end{align}

These expressions generalize the ones given in \cite{Sarlet1980}. Indeed if we
introduce the notation $\omega(t)\equiv\alpha(t)$, $\Phi\left(  t\right)
\equiv-\beta(t)$, then equations (\ref{eq.nonl6.1}), (\ref{eq.nonl6.2}) for
$c_{3}=0$ become eqs. (25), (26) of \cite{Sarlet1980}.

\section{Cosmological exact solutions}

\label{sec.exact.solution}

We can use the above results as an alternative to the Euler-Duarte-Moreira
method of integrability of the anharmonic oscillator
\cite{Duarte1991} in order to find exact solutions in the modified Brans-Dicke
(BD) theory.

Specifically, we consider the equation of motion for the quintessence scalar
field $\psi\left(  t\right)  $ with potential function $V(\psi)=\frac
{\psi^{n+1}}{n+1}$, where $n\neq-1$. Then equation (\ref{bd.14a})
becomes
\begin{equation}
\ddot{\psi}=-\psi^{n}-3\frac{\dot{a}}{a}\dot{\psi}\label{eq.nonl8}%
\end{equation}
which is a subcase of (\ref{eq.nonl1}) for $\omega(t)=1$ and $\Phi(t)=-3 \dot{(\ln
a)}$. Replacing in the transformation (\ref{eq.damp0c}) we find that
\begin{equation}
\tau(t)=\int a^{-3}(t)dt,\enskip\bar{\omega}(\tau(t))=a^{6}(t).\label{new0}%
\end{equation}
where equation (\ref{eq.nonl8}) now reads%
\begin{equation}
\psi^{\prime\prime}+a^{6}\psi^{n}=0\label{eq.01}%
\end{equation}
where $\psi^{\prime}=\frac{d\psi\left(  \tau\right)  }{d\tau}$.

The latter transformation for the background space becomes%
\begin{equation}
ds^{2}=-a^{6}\left(  \tau\right)  d\tau^{2}+a^{2}\left(  \tau\right)  \left(
dx^{2}+dy^{2}+dz^{2}\right)  \label{eq.02}%
\end{equation}
which means that the rest of the field equations read
\begin{align}
6\phi\left(  \frac{a^{\prime}}{a}\right)  ^{2} +6 \frac{a^{\prime}}{a}
\phi^{\prime} -\omega_{BD}\frac{\phi^{\prime2}}{\phi} -(\psi^{\prime})^{2}
-\frac{2}{n+1}a^{6}\psi^{n+1}  &  = 2a^{6}\rho_{m}\label{bd.17a}\\
4\phi\frac{a^{\prime\prime}}{a} -10\phi\left(  \frac{a^{\prime}}{a}\right)
^{2} -2 \frac{a^{\prime}}{a} \phi^{\prime} +\omega_{BD}\frac{\left(
\phi^{\prime}\right)  ^{2}}{\phi} +2\phi^{\prime\prime} +(\psi^{\prime})^{2}
-\frac{2}{n+1}a^{6}\psi^{n+1}  &  = -2a^{6}p_{m}\label{bd.17b}\\
6\phi\frac{a^{\prime\prime}}{a} -\omega_{BD}\left[  2\phi^{\prime\prime}%
-\frac{(\phi^{\prime})^{2}}{\phi}\right]  -12\phi\left(  \frac{a^{\prime}}%
{a}\right)  ^{2}  &  =0. \label{bd.17c}%
\end{align}

We proceed our analysis by constructing conservation laws for equation
(\ref{eq.01}) using the analysis presented in the previous section
\ref{sec.nonlin}.

\subsection{Case $n=0$}

For $n=0$ the associated QFI (\ref{eq.nonl4.1}) becomes
\begin{equation}
I=K_{11}\left(  \psi^{\prime}\right)  ^{2} -K_{11}^{\prime} \psi\psi^{\prime}
+b_{1}(\tau)\psi^{\prime} +c_{3}\psi^{2} +2a^{6}K_{11}\psi-b_{1}^{\prime}%
\psi+\int b_{1}(\tau) a^{6} d\tau\label{new1}%
\end{equation}
where $K_{11}=c_{1}+c_{2}\tau+c_{3}\tau^{2}$, the parameters $c_{1}%
,c_{2},c_{3}$ are arbitrary constants and the functions $b_{1}(\tau), a(\tau)$
satisfy the condition
\begin{equation}
b_{1}^{\prime\prime} =12a^{5}a^{\prime} K_{11} +3a^{6}K_{11}^{\prime}.
\label{new2}%
\end{equation}

We note that for $b_{1}=0$ we find the results of the subsection
\ref{sec.case4} below when $n=0$.

\subsection{Case $n=1$}

Using the transformation (\ref{new0}) equation $\psi^{\prime\prime}=-a^{6}\psi$
admits the solution
\begin{equation}
\psi(\tau)= \rho(\tau) \left(  A \sin\theta+B\cos\theta\right)  \label{new3}%
\end{equation}
where $\theta= \int\rho^{-2}d\tau$ and the functions $\rho(t(\tau)),
a(t(\tau))$ satisfy the condition
\begin{equation}
\rho^{\prime\prime} +\rho a^{6} -\rho^{-3}=0. \label{new4}%
\end{equation}

\subsection{Case $n=2$}

For $n=2$ we have $K_{11}= a^{-12/5}$ and the associated QFI (\ref{eq.nonl4.3}%
) becomes
\begin{small}
\begin{equation}
I=a^{-12/5} \left(  \psi^{\prime}\right)  ^{2} +\frac{12}{5} a^{-17/5}
a^{\prime} \psi\psi^{\prime} +(c_{4}+c_{5}\tau) \psi^{\prime} +\frac{2}{3}
a^{18/5} \psi^{3} +\frac{6}{5} \left[  \frac{17}{5} a^{-22/5} (a^{\prime})^{2}
- a^{-17/5} a^{\prime\prime}\right]  \psi^{2} -c_{5}\psi\label{new5}%
\end{equation}
\end{small}
where $c_{4},c_{5}$ are arbitrary constants and the function $a(t(\tau))\equiv
a(\tau)$ is given by
\begin{equation}
a^{\prime\prime\prime}-\frac{51}{5} \frac{a^{\prime}}{a}a^{\prime\prime}%
+\frac{374}{25} \left(  \frac{a^{\prime}}{a} \right)  ^{2}a^{\prime} +\frac
{5}{6} (c_{4}+c_{5}\tau) a^{47/5}=0. \label{eq.nonl4.4}%
\end{equation}

Substituting the given functions $\omega(t),\Phi(t)$ in equations
(\ref{eq.nonl4.4.1}) - (\ref{eq.nonl4.4.3}) we find equivalently that
\begin{equation}
a(t)=K_{11}^{-\frac{5}{12}} \label{eq.nonl8.0.6}%
\end{equation}%
\begin{align}
I  &  =K_{11}^{-3/2}\dot{\psi}^{2}-K_{11}^{-5/2}\dot{K}_{11}\psi\dot{\psi
}+\left(  c_{4}+c_{5}\int K_{11}^{5/4}dt\right)  K_{11}^{-5/4}\dot{\psi}%
+\frac{2}{3}K_{11}^{-3/2}\psi^{3} \nonumber\\
&  +\left[  \ddot{K}_{11}-\frac{5}{4}\left(  \ln K_{11}\right)  ^{\cdot}%
\dot{K}_{11}\right]  K_{11}^{-5/2}\frac{\psi^{2}}{2}-c_{5}\psi.
\label{eq.nonl8.0.7}%
\end{align}
where the function $K_{11}=K_{11}(t)$ is given by the differential equation
\begin{equation}
\dddot{K}_{11}-\frac{15}{4}\left(  \ln K_{11}\right)  ^{\cdot}\ddot{K}%
_{11}-\frac{5}{4}\left(  \ln K_{11}\right)  ^{\cdot\cdot}\dot{K}_{11}%
+\frac{25}{8}\frac{\dot{K}_{11}^{3}}{K_{11}^{2}}=2\left[  c_{4}+c_{5}\int
K_{11}^{5/4}dt\right]  K_{11}^{5/4}. \label{eq.nonl8.0.8}%
\end{equation}

Equation (\ref{eq.nonl8}) becomes $\ddot{\psi}=-\psi^{2}+\frac{5}{4}\left(
\ln K_{11}\right)  ^{\cdot}\dot{\psi}$. We note that for $c_{4}=c_{5}=0$ we
retrieve the results of the subsection \ref{sec.case4} below for $n=2$.

In the special case with $c_{5}=0$, we find for equation (\ref{eq.nonl8.0.8})
the special solution $K_{11}\left(  t\right)  =k_{0}t^{-12}$ with constraint
$c_{4}k_{0}^{1/4}=-192$ where $k_{0}$ is an arbitrary constant. Moreover from
equation (\ref{eq.nonl8.0.6}) the scale factor is determined
\begin{equation}
a(t)=K_{11}^{-\frac{5}{12}}=k_{0}^{-5/12}t^{5}.
\end{equation}
Therefore the Klein-Gordon equation (\ref{eq.nonl8}) becomes
\begin{equation}
\ddot{\psi}+\frac{15}{t}\dot{\psi}+\psi^{2}=0. \label{eq.KG1}%
\end{equation}

The latter equation can be solved by quadratures. In particular admits the Lie
symmetries
\[
\Gamma^{1}=\psi\partial_{\psi}-\frac{1}{2}t\partial_{t}~,~\Gamma^{2}=\left(
3\psi t^{2}-48\right)  \partial_{\psi}-\frac{1}{2}t^{3}\partial_{t}.
\]
By using the vector field $\Gamma^{1}$\ we find the reduced equation $\frac
{1}{2}\frac{d}{d\lambda}f^{2}+2\lambda\frac{d}{d\lambda}f+12f+\lambda^{2}%
=0$\ in which $f\left(  \lambda\right)  =t^{3}\dot{\psi}~,~\lambda=t^{2}\psi$.
The latter equation is an Abel equation of second type. Moreover if we assume
that $\lambda$\ is a constant, $\lambda=\lambda_{0}$\ \ then we find
$\psi=\lambda_{0}t^{-2}$\ where by replacing in equation (\ref{eq.KG1}) it
follows $\lambda_{0}=24$. Therefore we end up with the solution $\psi=
\frac{24}{t^{2}}$. Let us now find the complete solution for the gravitational
field equations for this particular exact solution.

Replacing these results in the rest of the field equations for dust fluid
source, that is, $p_{m}=0$ and $\rho_{m}=\rho_{0} a^{-3}$ where $\rho_{0}$ is
a constant, the evolution equation for the Brans-Dicke field becomes
\[
\ddot{\phi} +\frac{15}{t}\dot{\phi} = \frac{1}{2\omega+3} \left(  \rho_{0}
a^{-3} -\dot{\psi}^{2} +\frac{4}{3}\psi^{3} \right)
\]
which admits the general solution
\[
\phi(t)= -\frac{1}{2\omega+3} \left(  \frac{2016}{5} t^{-4} +\frac{\rho
_{0}k_{0}^{5/4}}{13}t^{-13} \right)  +\frac{k_{1}}{14}t^{-14}
\]
where $k_{1}$ is an arbitrary constant. Finally by replacing in the constraint
equation (\ref{bd.10}) follows (eq. (\ref{bd.11}) is satisfied identically)
\[
\omega=-\frac{45}{16}, \enskip k_{1}=\rho_{0}=0.
\]

We conclude that the gravitational field equations for this model with the
use of the QFI for equation (\ref{eq.nonl8}) admit the following exact solution 
\begin{equation}
\omega=-\frac{45}{16}, \enskip a(t)= k_{0}^{-5/12} t^{5}, \enskip \psi
(t)=24t^{-2}, \enskip \phi(t)= \frac{768}{5}t^{-4} \label{sol1}%
\end{equation}
with physical quantities
\[
\rho_{m}=p_{m}=0, \enskip \rho_{\psi}= 5760 t^{-6}, \enskip p_{\psi}= -3456
t^{-6}.
\]

For the solution (\ref{sol1}) the transformation (\ref{new0}) gives
\begin{equation}
\tau= -\frac{k_{0}^{5/4}}{14}t^{-14} \implies t= \left(  -14 k_{0}^{-5/4}
\right)  ^{-1/14} \tau^{-1/14}. \label{sol1.1}%
\end{equation}
Then the transformed field equations (\ref{eq.01}) and (\ref{bd.17a}) -
(\ref{bd.17c}) admit the solutions
\begin{align}
&\omega=-\frac{45}{16}, \enskip a= k_{0}^{-5/12} (-14 k_{0}^{-5/4})^{-5/14}%
\tau^{-5/14}, \enskip \nonumber \\
& \psi=24(-14 k_{0}^{-5/4})^{1/7}\tau^{1/7}, \enskip \phi=
\frac{768}{5}(-14 k_{0}^{-5/4})^{2/7}\tau^{2/7}. \label{sol3}%
\end{align}

\subsection{Case $n\neq-1$}

\label{sec.case4}

In this case the associated QFI (\ref{eq.nonl5}) becomes
\begin{equation}
I=(c_{1}+c_{2}\tau+c_{3}\tau^{2})\left(  \psi^{\prime} \right)  ^{2}%
-(c_{2}+2c_{3}\tau) \psi\psi^{\prime} +\frac{2}{n+1}(c_{1} +c_{2}\tau
+c_{3}\tau^{2})^{-\frac{n+1}{2}}\psi^{n+1} +c_{3}\psi^{2} \label{new6}%
\end{equation}
and the function $a(\tau)$ is
\begin{equation}
a(\tau) =(c_{1}+c_{2}\tau+c_{3}\tau^{2})^{-\frac{n+3}{12}}. \label{new7}%
\end{equation}

Substituting the given functions $\omega(t), \Phi(t)$ in the relation
(\ref{eq.nonl6.1}) we find equivalently that
\begin{equation}
a^{6}(t)= \left[  c_{1} +c_{2}\int a^{-3}(t) dt +c_{3}\left(  \int a^{-3}(t)
dt\right)  ^{2} \right]  ^{-\frac{n+3}{2}} \label{eq.nonl8.1}%
\end{equation}
and the associated QFI (\ref{eq.nonl6.2}) becomes
\begin{align}
I  &  = \left[  c_{1} +c_{2}\int a^{-3}(t) dt +c_{3}\left(  \int a^{-3}(t)
dt\right)  ^{2} \right]  a^{6}(t) \dot{\psi}^{2} \nonumber \\
& -\left[  c_{2} +2c_{3}\int
a^{-3}(t) dt \right]  a^{3}(t) \psi\dot{\psi}  \nonumber\\
&  +\frac{2}{n+1} \left[  c_{1} +c_{2}\int a^{-3}(t) dt +c_{3}\left(  \int
a^{-3}(t) dt\right)  ^{2} \right]  ^{-\frac{n+1}{2}} \psi^{n+1} + c_{3}%
\psi^{2}. \label{eq.nonl8.2}%
\end{align}

We consider the following special cases for which equation (\ref{eq.nonl8})
admits a closed-form solution for $n\neq-3,1$. In the case $n=-3$ the spacetime
is that of Minkowski  space. Hence we omit the analysis.

\subsubsection{Subcase $\left\vert \tau\right\vert <<1$}

For small values of $\left\vert \tau\right\vert $ (i.e. $c_{1}=c_{3}=0$) the scale
factor (\ref{new7}) is approximated as $a\left(  \tau\right)  \simeq
\tau^{-\frac{n+3}{12}}$, therefore it follows
\begin{equation}
a(t)=B_{0} (t-t_{0})^{\frac{n+3}{3(n-1)}} \label{eq.nonl9}%
\end{equation}
where $B_{0}= \left[  -\frac{c_{2}(n-1)}{4} \right]  ^{\frac{n+3}{3(n-1)}}$
and $t_{0}$ is an arbitrary constant.

For this asymptotic solution the equation of motion (\ref{eq.nonl8}) for the
second field $\psi$ becomes
\begin{equation}
\ddot{\psi}=-\psi^{n}-\frac{n+3}{n-1}\frac{1}{t-t_{0}} \dot{\psi}.
\label{eq.nonl9.1}%
\end{equation}
For the latter equation the QFI (\ref{eq.nonl8.2}) is
\begin{equation}
I=\left[  -\frac{c_{2}(n-1)}{4}\right]  ^{\frac{2(n+1)}{n-1}}(t-t_{0}%
)^{\frac{2(n+1)}{n-1}}\left(  \dot{\psi}^{2}+\frac{2}{n+1}\psi^{n+1}\right)
-c_{2}\left[  -\frac{c_{2}(n-1)}{4}\right]  ^{\frac{n+3}{n-1}}(t-t_{0}%
)^{\frac{n+3}{n-1}}\psi\dot{\psi}. \label{eq.nonl10}%
\end{equation}
This QFI for the scale factor (\ref{eq.nonl9}) together with the results of
the cases $n=0,1,2$ produce new solutions $\psi(t)$ which  have not found before.

Furthermore, for the scale factor (\ref{eq.nonl9}) the closed-form solution for the scalar field $\psi\left(  t\right)  $ from (\ref{eq.nonl9.1}) is derived
\begin{equation}
\psi\left(  t\right)  =\psi_{0}(t-t_{0})^{-\frac{2}{n-1}}~,~\psi_{0}=\left(
\frac{2}{n-1}\right)  ^{\frac{2}{n-1}} \label{eq.nonl8.6.1}%
\end{equation}
whereas for the BD field $\phi\left(  t\right)  $ it follows that $n=3,~\phi(t)=\frac{\phi_{0}}{(t-t_{0})^{2}}$
and~$~\omega_{BD}=-\frac{3}{2}$. However, this value for the BD parameter
$\omega_{BD}$ is not physically acceptable. Hence we do not have any close-form
solution. In all discussion above we have considered $\rho_{m}=0$.

The closed-form solution found in this section is not the general solution of
the field equations. That is easy to be seen since they have less free
parameters from the degrees of freedom of the dynamical system. However, this
form of solutions are of special interest in cosmological studies because they
can describe various phases of the cosmological evolution, such as  the early
inflationary epoch.

\subsubsection{Subcase $\left\vert \tau\right\vert \gg1$}

For large values of $\tau\gg0$ (i.e. $c_{1}=c_{2}=0$), the scale factor
(\ref{new7}) is approximated as $a(\tau)\simeq\tau^{-\frac{n+3}{6}}$.
Therefore, in the original variable equation (\ref{eq.nonl8.1}) becomes
\begin{equation}
a^{-\frac{6}{n+3}}=c_{3}^{\frac{1}{2}}\int a^{-3}dt \label{eq.nonl8.3}%
\end{equation}
which implies (see eq. (31) of \cite{Mukherjee2019})
\begin{equation}
a(t)=A_{0}(t-t_{0})^{\frac{n+3}{3(n+1)}} \label{eq.nonl8.4}%
\end{equation}
where $A_{0}=\left[  -\frac{\sqrt{c_{3}}(n+1)}{2}\right]  ^{\frac{n+3}%
{3(n+1)}}$ and $t_{0}$ is an arbitrary constant. The scale factor
(\ref{eq.nonl8.4}) describes a scaling solution where the effective
cosmological fluid is that of an ideal gas with effective parameter for the
equation of state $w_{eff}=\frac{n-1}{n+3}.$ Furthermore, for
$-3<n<-1~,~-1<n<0$ the scale factor describes an accelerated universe. For
$-1<n<0,~w_{eff}$ is bounded as $-1<w_{eff}<-\frac{1}{3}$ while for $-3<n<-1$,
$w_{eff}$ crosses the phantom divide line, that is $w_{eff}<-1$.

For this asymptotic solution the equation of motion (\ref{eq.nonl8}) for the
second field $\psi$ becomes%
\begin{equation}
\ddot{\psi}= -\psi^{n} -\frac{n+3}{n+1} \frac{1}{t-t_{0}} \dot{\psi}
\label{eq.nonl8.3.1}%
\end{equation}
and the corresponding QFI (\ref{eq.nonl8.2}) is written as
\begin{equation}
I= c_{3} \left[  \frac{(n+1)(t-t_{0})}{2} \dot{\psi} +\psi\right]  ^{2} +
\frac{c_{3}(n+1)}{2} (t-t_{0})^{2} \psi^{n+1} \label{eq.nonl8.5}%
\end{equation}
where $t\neq t_{0}$.

However, the system admits the closed form solution (see eq. (32) of
\cite{Mukherjee2019})
\begin{equation}
\psi\left(  t\right)  =\psi_{0}\left(  t-t_{0}\right)  ^{-\frac{2}{n-1}%
}\label{eq.nonl8.5.1}%
\end{equation}
in which $\psi_{0}$ is given by the expression $\psi_{0}=(-2)^{\frac{3}{n-1}%
}\left[  \left(  n+1\right)  \left(  n-1\right)  ^{2}\right]  ^{\frac{1}{1-n}%
}$. Replacing in the remaining equations (\ref{bd.10}) - (\ref{bd.13}) for the
Brans-Dicke field we calculate%
\begin{equation}
\phi\left(  t\right)  =\phi_{0}(t-t_{0})^{-\frac{4}{n-1}}\label{eq.nonl8.5.2}%
\end{equation}
in which
\begin{align}
&\phi_{0}   =\frac{(n-1)^{\frac{4}{1-n}}}{2(n+3)(2\omega_{BD}+3)}\left[
(-2)^{\frac{3(n+1)}{n-1}}(n+1)^{\frac{n+1}{1-n}}-(-2)^{\frac{6}{n-1}%
}(n+1)^{\frac{n-3}{n-1}}\right]  \label{eq.nonl8.5.3}\\
 & \omega_{BD} =\frac{b_{1}-3b_{2}}{1+2b_{2}}\label{eq.nonl8.5.4}%
\end{align}
while we have assumed that there is not any other matter source, i.e.
$\rho_{m}=0$. The constants $b_{1},b_{2}$ are given by the relations
\begin{align}
b_{1} &  =\frac{(n+3)(n-1)}{2(n+1)}\left[  \frac{(n+3)(n-1)}{12(n+1)}%
-1\right]  \label{eq.nonl8.5.5}\\
b_{2} &  =\frac{n+3}{4}\cdot\frac{2(-2)^{\frac{6}{n-1}}(n+1)^{\frac{2}{1-n}%
}+(-2)^{\frac{3(n+1)}{n-1}}(n+1)^{\frac{2n}{1-n}}}{(-2)^{\frac{3(n+1)}{n-1}%
}(n+1)^{\frac{n+1}{1-n}}-(-2)^{\frac{6}{n-1}}(n+1)^{\frac{n-3}{n-1}}%
}.\label{eq.nonl8.5.6}%
\end{align}

In the following we perform a detailed study on the stability of the latter
closed-form solutions.

\section{Stability of scaling solutions}
\label{sec.stab}

According to the methods in \cite{Ratra:1987rm,Liddle:1998xm,Uzan:1999ch} let be
\begin{equation}
    F(\ddot\psi, \dot\psi, \psi)=0
\end{equation}
a second order ODE in one dimension which admits a singular power law solution
\begin{equation}
    \psi_c(t)= \psi_0 t^\beta
\end{equation}
where $\psi_{0}$ is an arbitrary constant. To examine the stability of the solution $\psi_c$, the logarithmic time $T$ through
$t= e^{T}$ is introduced, such that $t\rightarrow 0$ as  $T \rightarrow -\infty$ and $t\rightarrow +\infty$ as $T\rightarrow +\infty$. We use $\psi'\equiv \frac{d \psi}{dT}$ in the following discussion.

The following dimensionless function is introduced\begin{equation}
    u(T)= \frac{\psi(T)}{\psi_c(T)}
\end{equation}
and the stability analysis in translated into the analysis of the stability of the equilibrium point $u=1$ of a transformed dynamical system.
To construct the aforementioned system the following relations are useful:
\begin{equation}
\label{eq.82}
  \dot\psi= e^{-T}\psi',  \quad  \ddot\psi= e^{-2 T} (\psi''-\psi'), \quad \text{and} \quad
    \frac{\psi_c'}{\psi_c}=\beta \quad \text{if} \quad \psi_c(t)= \psi_0 t^\beta.
\end{equation}
In this section we use a similar procedure for analyzing  stability of the scaling solutions obtained in section \ref{sec.case4}.

\subsection{Case $\left\vert \tau\right\vert \gg 1$}

For the analysis of the solution \eqref{eq.nonl8.5.1}  of  \eqref{eq.nonl8.3.1} we set $t_0=0$ by a time shift. Using \eqref{eq.82} we have
\begin{equation}
\label{eq.83}
    \psi ''(T )= -\frac{2 \psi '(T )}{n+1}-e^{2 T } \psi (T )^n.
\end{equation}
Denoting $p= -\frac{2}{n-1}$ we have
\begin{align}
& u''(T
   )= \frac{p^2 e^{-p T } \psi (T
   )}{\psi_0}+\frac{e^{-p T } \psi ''(T )}{\psi_0}-\frac{2 p e^{-p T } \psi '(T )}{\psi_0}\\
& u'(T )= \frac{e^{-p T } \psi '(T )}{\psi_0}-\frac{p e^{-p
   T } \psi (T )}{\psi_0}\\
& u(T )=\frac{e^{-p T } \psi
   (T )}{\psi_0}.
\end{align}
Hence
\begin{align}
& \psi ''(T )= \psi_0 e^{p T } \left(p^2
   u(T )+2 p u'(T )+u''(T )\right)\\
& \psi '(T )= \psi_0 e^{p T } \left(p u(T )+u'(T )\right)\\
& \psi (T ))= \psi_0 e^{p T } u(T ).
\end{align}
Equation \eqref{eq.83}
becomes
\begin{align}
    u''(T )= \left(-\frac{2}{n+1}-2 p\right) u'(T )+\psi_0^{n-1} \left(-e^{T  ((n-1) p+2)}\right) u(T )^n-\frac{p (n
   p+p+2) u(T )}{n+1}.
\end{align}
Substituting $p= -\frac{2}{n-1}$ and $\psi_{0}=(-2)^{\frac{3}{n-1}%
}\left[  \left(  n+1\right)  \left(  n-1\right)  ^{2}\right]  ^{\frac{1}{1-n}%
}$ it is obtained the second order equation
\begin{align}
    u''(T )=\frac{2 (n+3) u'(T )}{n^2-1}+\frac{8 u(T )^n}{(n-1)^2
   (n+1)}-\frac{8 u(T )}{(n-1)^2 (n+1)}.
\end{align}
Defining
\begin{equation}
    x= u(T), \quad y=u'(T)
\end{equation}
we obtain the autonomous system
\begin{align}
& x'(T)= y(T) \label{eq.93}\\
&    y'(T )=\frac{2 (n+3) y(T )}{n^2-1}+\frac{8 x(T
   )^n}{(n-1)^2 (n+1)}-\frac{8 x(T )}{(n-1)^2 (n+1)}. \label{eq.94}
\end{align}

\begin{figure*}
    \centering
    \includegraphics[width=0.75\textwidth]{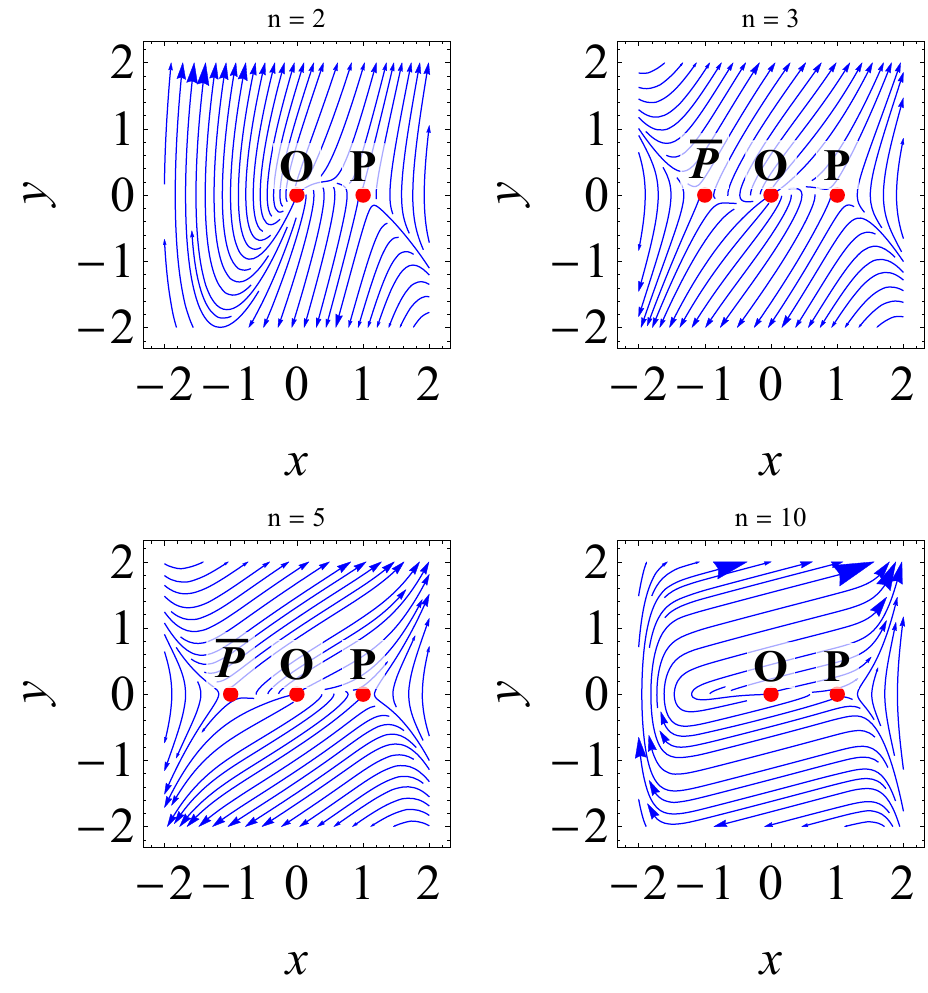}
    \caption{Phase-plot of system  \eqref{eq.93}, \eqref{eq.94} for $n\in\{2, 3, 5, 10\}$. $P$ is a saddle given $|n|>1$. When $n$ is odd, the symmetrical point $\bar{P}$ is a saddle given $|n|>1$. The origin $O$ is a source, and the orbits diverge to infinity.}
    \label{fig:1}
\end{figure*}

\begin{figure*}
    \centering
    \includegraphics[width=0.75\textwidth]{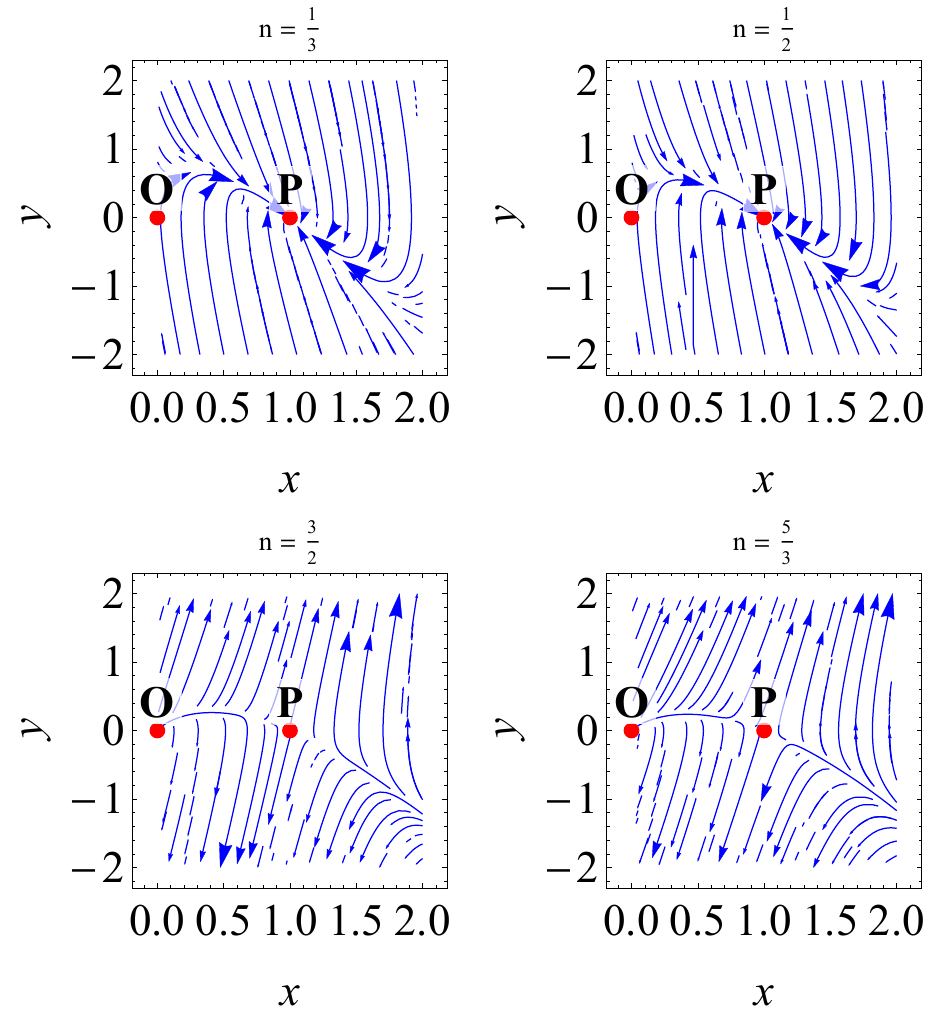}
    \caption{Phase-plot of system  \eqref{eq.93}, \eqref{eq.94} for $n\in\{2, 3, 5, 10\}$. $P$ is a saddle given $|n|>1$. When $n$ is odd, the symmetrical point $\bar{P}$ is a saddle given $|n|>1$.}
    \label{fig:2}
\end{figure*}

The scaling solution \eqref{eq.nonl8.5.1}  is transformed to the equilibrium point $P:= (x,y)=(1,0)$. The system \eqref{eq.93}, \eqref{eq.94} also admits the trivial solution  $O: (x,y)=(0,0)$ as an equilibrium point and in case that $n$ is odd, the symmetrical point $P$ given by $\bar{P}:= (x,y)=(-1,0)$ is also an equilibrium point.

The linearization matrix of system \eqref{eq.93}, \eqref{eq.94}  is
\begin{equation}
    J(x,y)=\left(
\begin{array}{cc}
 0 & 1 \\
 \frac{8 \left(n x^{n-1}-1\right)}{(n-1)^2 (n+1)} & \frac{2 (n+3)}{n^2-1}
   \\
\end{array}
\right).
\end{equation}
For $n>1$, $J(0,0)$ is real-valued, with eigenvalues $\left\{\frac{4}{n^2-1},\frac{2}{n-1}\right\}$. Then the origin is unstable for $n>1$.

The eigenvalues of $J(1,0)$ are $\left\{-\frac{2}{n+1},-\frac{4}{1-n}\right\}$. Therefore, $(x,y)=(1,0)$ is a sink for $-1<n<1$. It is a saddle for $n<-1$, or $n>1$.

If $n$ is odd number, say $n=2 k+1$, with $k \in \mathbb{Z}$, the eigenvalues of $J(-1,0)$ are $\left\{-\frac{1}{k+1},\frac{2}{k}\right\}$ and when it exists, $\bar{P}$ is a saddle.

In Figure \ref{fig:1} a phase-plot of system  \eqref{eq.93}, \eqref{eq.94} for  $n\in\{2, 3, 5, 10\}$ is presented. $P$ is a saddle given $|n|>1$. When $n$ is odd, the symmetrical point $\bar{P}$ is a saddle given $|n|>1$. The origin $O$ is a source, and the orbits diverge to infinity.

In Figure \ref{fig:2} a phase-plot of system  \eqref{eq.93}, \eqref{eq.94} for $n\in\{1/3, 1/2, 3/2, 5/2\}$ is presented. When $n<1$ the power law solution $P$ is a sink, whereas in the other cases is a saddle given $|n|>1$.

\section{Conclusions}
\label{sec.con}

In this work we considered a cosmological model consisted by a Brans-Dicke field and a minimally coupled quintessence field in a spatially flat FLRW background space. For this cosmological model the gravitational field equations consist a Hamiltonian system of six degrees of freedom. The dynamical variables correspond to the scale factor and to the two scalar fields.

In order to study the integrability of the field equations we have applied a direct method which determines the FIs of a dynamical system without the use of Noether's theorem. In this approach one assumes a generic form for the FIs, say $I$, and applies directly the condition $dI/dt=0$ using the dynamical
equations. These considerations resulted in a system of partial differential equations involving the unknown coefficients defining $I$ and the dynamical quantities which characterize the dynamical system. The resulting system of equations is solved in terms of the symmetries and the Killing tensors of the kinetic metric and its solution provides the considered FIs.

For a power law scalar field potential function of the quintessence field we found conservation laws quadratic in the first order derivatives. Using the conservation laws we were able to find exact solutions for the field equations. In particular we found scaling solutions for the scale factor which describe ideal gas solutions. The stability properties of these solutions was investigated. We were able to recover previous published results in the literature and also to find new QFIs.

Using  methods in \cite{Ratra:1987rm,Liddle:1998xm,Uzan:1999ch} we have studied second order ODE in one dimension which admits a singular power law solution $\psi_c(t)= \psi_0 t^\beta$
where $\psi_{0}$ is an arbitrary constant. To examine the stability of the solution $\psi_c$, the logarithmic time $T$ through
$t= e^{T}$ was introduced, such that $t\rightarrow 0$ as  $T \rightarrow -\infty$ and $t\rightarrow +\infty$ as $T\rightarrow +\infty$. According to our analysis, the scaling solution \eqref{eq.nonl8.5.1}  is transformed to the equilibrium point $P:= (x,y)=(1,0)$, which is a sink for $-1<n<1$ or a saddle for $n<-1$, or $n>1$. The dynamical system also admits the trivial solution  $O: (x,y)=(0,0)$ as an equilibrium point and in case that $n$ is odd, the symmetrical point $P$ given by $\bar{P}:= (x,y)=(-1,0)$ is also an equilibrium point. The origin is unstable for $n>1$. If $n$ is odd number, the point $\bar{P}$  exists and it is a saddle. 

Until now, the majority of this kind of studies, for the investigation of conservation laws, have been done mainly with the application of variational symmetries. Our approach is more general and does not required the existence of a point-like Lagrangian, that is, of a minisuperspace description. Therefore, this generic approach can be applied in other gravitational models without minisuperspace such are the Class B Bianchi spacetimes.

\begin{acknowledgments}
The research of AP and GL was funded by Agencia Nacional de Investigaci\'{o}n
y Desarrollo - ANID through the program FONDECYT Iniciaci\'{o}n grant no.
11180126. Additionally, GL was funded by Vicerrector\'{\i}a de
Investigaci\'{o}n y Desarrollo Tecnol\'{o}gico at Universidad Cat\'olica del
Norte. This work is based on the research supported in part by the National
Research Foundation of South Africa (Grant Numbers 131604).
\end{acknowledgments}

\end{document}